\documentclass[aps,prl,superscriptaddress,two column,preprintnumbers,notitlepage]{revtex4-1}
\usepackage{latexsym}
\usepackage{subcaption}
\usepackage{caption}

\usepackage[dvips]{color}
\usepackage{graphicx}
\usepackage{amsmath}
\usepackage{amssymb}
\usepackage{gensymb}
\usepackage{enumerate}
\usepackage{hyperref}
\usepackage{mathrsfs}
\usepackage[usenames,dvipsnames]{xcolor}
\usepackage[normalem]{ulem}
\usepackage{float}
\def\mr{\mathrm}

\begin{document}

\title{Quantum Hall interferometry in triangular domains of marginally twisted bilayer graphene}

\author{Phanibhusan S. Mahapatra}
\email{phanis@iisc.ac.in}
\affiliation{Department of Physics, Indian Institute of Science, Bangalore, 560012, India}
\author{Manjari Garg}
\affiliation{Department of Instrumentation and Applied Physics, Indian Institute of Science, Bangalore, 560012, India}
\author{Bhaskar Ghawri}
\affiliation{Department of Physics, Indian Institute of Science, Bangalore, 560012, India}
\author{Aditya Jayaraman}
\affiliation{Department of Physics, Indian Institute of Science, Bangalore, 560012, India}

\author{Kenji Watanabe}
\affiliation{Research Center for Functional Materials, National Institute for Materials Science, Namiki 1-1, Tsukuba, Ibaraki 305-0044, Japan}
\author{ Takashi Taniguchi}
\affiliation{International Center for Materials Nanoarchitectonics, National Institute for Materials Science, Namiki 1-1, Tsukuba, Ibaraki 305-0044, Japan}

\author{Arindam Ghosh}
\affiliation{Department of Physics, Indian Institute of Science, Bangalore, 560012, India}
\affiliation{Centre for Nano Science and Engineering, Indian Institute of Science, Bangalore 560 012, India}
\author{U. Chandni}
\affiliation{Department of Instrumentation and Applied Physics, Indian Institute of Science, Bangalore, 560012, India}


\begin{abstract}
Quantum Hall (QH) interferometry provides an archetypal platform for the experimental realization of braiding statistics of fractional QH states. However, the complexity of observing fractional statistics requires phase coherence over the length of the interferometer,  as well as suppression of Coulomb charging energy. Here, we demonstrate a new type of QH interferometer based on marginally twisted bilayer graphene (mtBLG), with a twist angle $\theta \approx 0.16\degree$. With the device operating in the QH regime, we observe distinct signatures of electronic Fabry-P\'{e}rot (FP) and Aharonov-Bohm (AhB)-oscillations of the magneto-thermopower in the density-magnetic field phase-space, at Landau level filling factors $\nu=4$,~$8$. We find that QH interference effects are intrinsic to the triangular AB/BA domains in mtBLG that show diminished Coulomb charging effects. Our results demonstrate phase-coherent interference of QH edge modes without any additional gate-defined complex architecture, which may be beneficial in experimental realizations of non-Abelian braiding statistics.
\end{abstract}

\maketitle
\section{Introduction}

The interference between edge modes in integer and fractional quantum Hall (QH) regimes can be used to directly probe the underlying hierarchy of the quasiparticle excitations \citep{nakamura2020direct,chamon1997two,nakamura2019aharonov,sarma2005topologically,stern2006proposed,kim2006aharanov,camino2005aharonov,lin2009electron,halperin1984statistics}. Intriguingly, this proposal can be particularly effective in observing the non-Abelian braiding statistics of quasiparticles in the fractional QH regime \cite{sarma2005topologically,nakamura2019aharonov}. The difficulty in experimental realizations of such phenomena involves the inevitable Coulomb repulsion from the confined quasiparticles, which changes the effective area of the interferometer \citep{halperin2011theory,von2015enhanced}. This additional charging effect is detrimental to observing robust braiding statistics, which can be somewhat mitigated by larger device dimensions, albeit at the expense of phase coherence between the interfering paths \cite{gurman2016dephasing}. This has led to various device architectures \cite{ofek2010role,bhattacharyya2019melting,ronen2021aharonov,willett2013magnetic}, and materials engineering \cite{nakamura2019aharonov} to suppress the Coulomb repulsion, while maintaining the phase coherence of the quasiparticles. 

Here, we report the operation of a new type of QH interferometer based on a mtBLG device at $\theta \approx 0.16\degree$. The moir\'{e} lattice and the corresponding band structure of tBLG at or near magic angle ($\theta_\mr{m} =1.1\degree$) can host a myriad of novel phases such as correlated insulators \citep{cao2018correlated}, superconductivity \citep{cao2018unconventional} and magnetism \citep{sharpe2019emergent}. When the twist angle is well below $\theta_\mr{m}$, relaxation effects change the atomic registries of the moir\'{e} lattice \citep{yoo2019atomic}. This leads to a mosaic structure of triangular regions consisting of alternate AB and BA stacking, which are separated by domain walls (Fig.~1a). When a vertical displacement field gaps out the AB/BA regions, the domains walls can host topologically-protected helical one dimensional ($1$D) network of conducting channels \cite{huang2018topologically,san2013helical,efimkin2018helical, walet2019emergence}. The quasiparticle transport through the helical $1$D network can support AhB-oscillations when subjected to a perpendicular magnetic field \cite{xu2019giant}. However, in previous reports~\cite{rickhaus2018transport,xu2019giant} the helical $1$D network remained insensitive to the QH phenomenon since the bulk AB/BA regions were depleted of charge carriers.

In this letter, we have performed electrical and thermoelectric measurements in mtBLG with $\theta \approx 0.16\degree$ in the integer QH regime at negligible displacement fields. Thermoelectric coefficient provides a fundamental characterization of the electronic state since the diffusive nature of the transport is characteristically sensitive to the scattering dynamics of the charge carriers \cite{mahapatra2020misorientation,izawa2007thermoelectric}.  Moreover, thermopower can be used to probe the additional degrees of freedom of the electronic state which are often undetectable in standard resistance measurements \cite{wang2003spin,scheibner2005thermopower}. Under appropriate conditions, the statistical properties of the non-Abelian quasiparticles can also be explored with thermoelectric measurements since the entropy of the anyons are larger compared to their Abelian counterparts \cite{yang2009thermopower}. Here, we use the dependence of thermal voltage on magnetic field ($B$) and gate-induced density ($n$) to probe the diverse characteristics of interference effects in the QH regime.

\begin{figure*}[t]
  \includegraphics[width=1.0\textwidth]{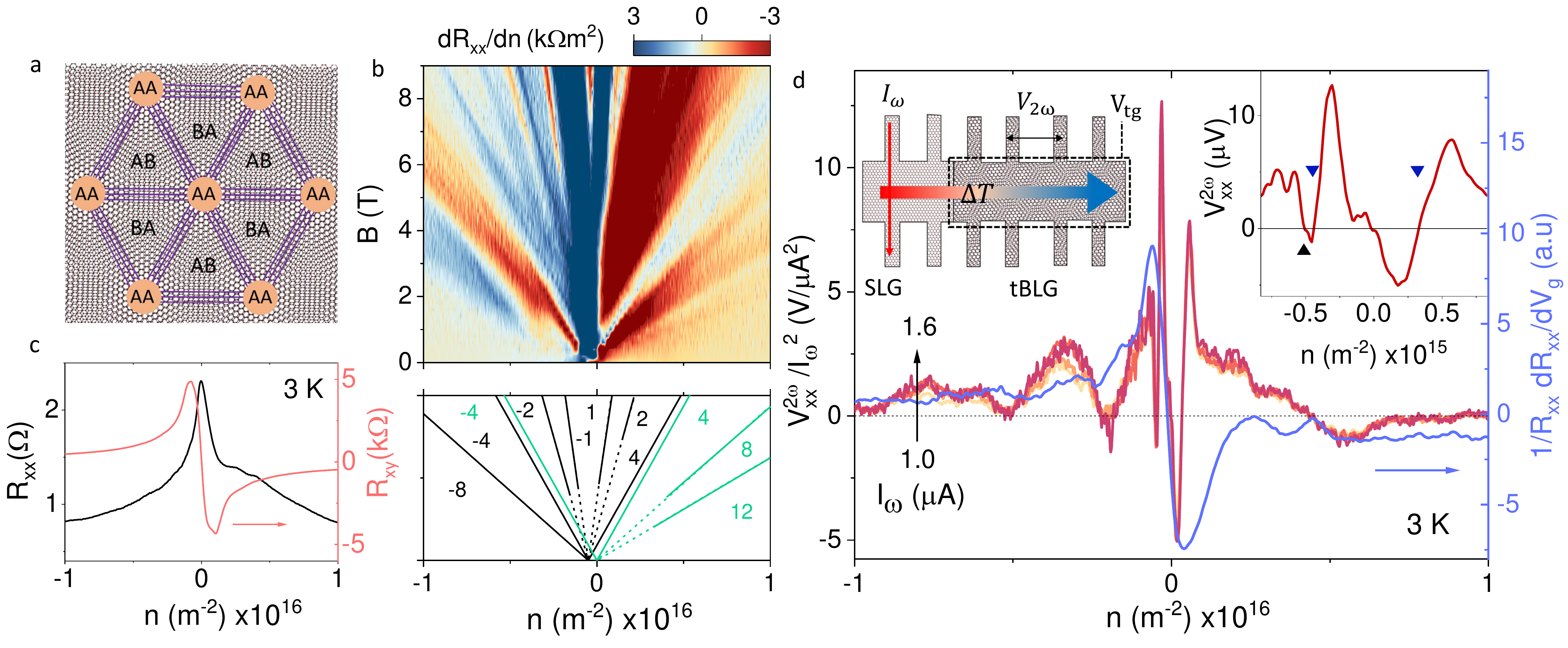}
  \captionsetup{justification=raggedright,singlelinecheck=false}
\caption{\textbf{Magneto-resistance and thermopower measurements:} (a) The AB/BA triangular domains in mtBLG separated by AA-stacked regions and domain walls. (b) Landau fan diagram of four terminal resistance ($R_\mr{xx}$). Measurements are taken at $3$~K. The bottom panel shows the reconstructed Landau level structure with filling factors ($\nu$). The green lines show the Landau fan emanating from the CNP, while the black lines show the secondary Landau fans emanating from the full-filling of the first moir\'{e} band on the hole-doping side. (c) Doping dependence of $R_\mr{xx}$ at $B=0$ and Hall resistance ($R_\mr{xy}$) at $800$~mT. (d) Doping dependence of Seebeck voltage $V_\mr{xx}^\mr{2\omega}$ normalized with $I_\mr{2\omega}^2$ for a range of $I_\mr{2\omega}$ at $3$~K. The blue curve shows $\gamma=1/R_\mr{xx}dR_\mr{xx}/dV_\mr{tg}$ for comparison on the right axis. The left inset illustrates the measurement schematic for thermopower. The right inset shows the magnified part of $V_\mr{xx}^\mr{2\omega}$ near CNP, where blue markers show the van Hove singularities (vHs) and the full-filling of the first moir\'{e} band (black marker). } 
\end{figure*}
\section{Results and Discussion}

The low-angle tBLG device was created using standard van der Waals stacking \citep{mahapatra2020misorientation}, consisting of two graphene layers aligned at $\theta$, where $\theta$ is the effective twist angle, and encapsulated within hexagonal boron nitride (hBN) layers. The device was further shaped into a Hall-bar, where the local metal top-gate tunes the charge carrier density ($n$) in the tBLG region, while the extended part of the single-layer graphene part outside the overlap region is used as the heater to set up a temperature gradient (see left inset of Fig.~1d). We perform the electrical and thermoelectric measurements at $3$~K and at negligible vertical displacement fields ($<0.1$~Vnm$^{-1}$) which are not sufficient to induce a band gap in the AB/BA regions. Fig.~1b shows the Landau fan diagram of the numerical derivative of the longitudinal resistance $dR_\mr{xx}/dn$ at $3$~K. The Landau fans show resemblance to the Hofstadter butterfly pattern, where the primary Landau levels (LL) emerge from the charge neutrality point (CNP) (green lines) at $n=0$ while on the hole-doped side, another set of LLs emerge (black dashed lines) from $n_\mr{s}=5 \times 10^{14}$~m$^{-2}$ \cite{dean2013hofstadter}. The secondary LLs account for the filling of the first mini band of the moir\'{e} lattice. We derive the moir\'{e} period $\lambda \approx 92$~nm from $n_\mr{s}$, with a corresponding twist angle $\theta \approx 0.16\degree$. The primary LLs originating from the CNP exhibit the filling factor sequence $\nu = \pm 4$, $\pm 8$, $\pm 12$ ($\nu=n\Phi_0/B$, where $\Phi_0=h/e$ is the magnetic quantum flux). This is similar to the four-fold degeneracy of spin and valley quantum numbers in monolayer graphene \cite{zhang2005experimental}. In contrast, the secondary LLs exhibit the distinct QH sequence $\nu=\pm 1$, $\pm 2$ for $B \gtrsim 5$~T. However, the sequence coincides with the four-fold degeneracy of $\nu = \pm 4$, $\pm 8$ at lower magnetic fields. The reduced degeneracy of the QH states at higher magnetic fields clearly indicates the breaking of spin/valley symmetry, in contrast to the quantum oscillations arising from the CNP. At $B=0$, however, the first mini-band shows no distinct peak in $R_\mr{xx}$, which has a pronounced maxima only at CNP (Fig.~1c). The absence of resistance maxima near $\pm n_\mr{s}$ is possibly due to the presence of a large number of overlapping bands at low energies~\cite{yoo2019atomic}. This is further supported by the density dependence of $R_\mr{xy}$ at $800$~mT (right axis of Fig.~1c) which shows the sign reversal only at CNP. 

\begin{figure*}[t]
  \includegraphics[width=1.0\textwidth]{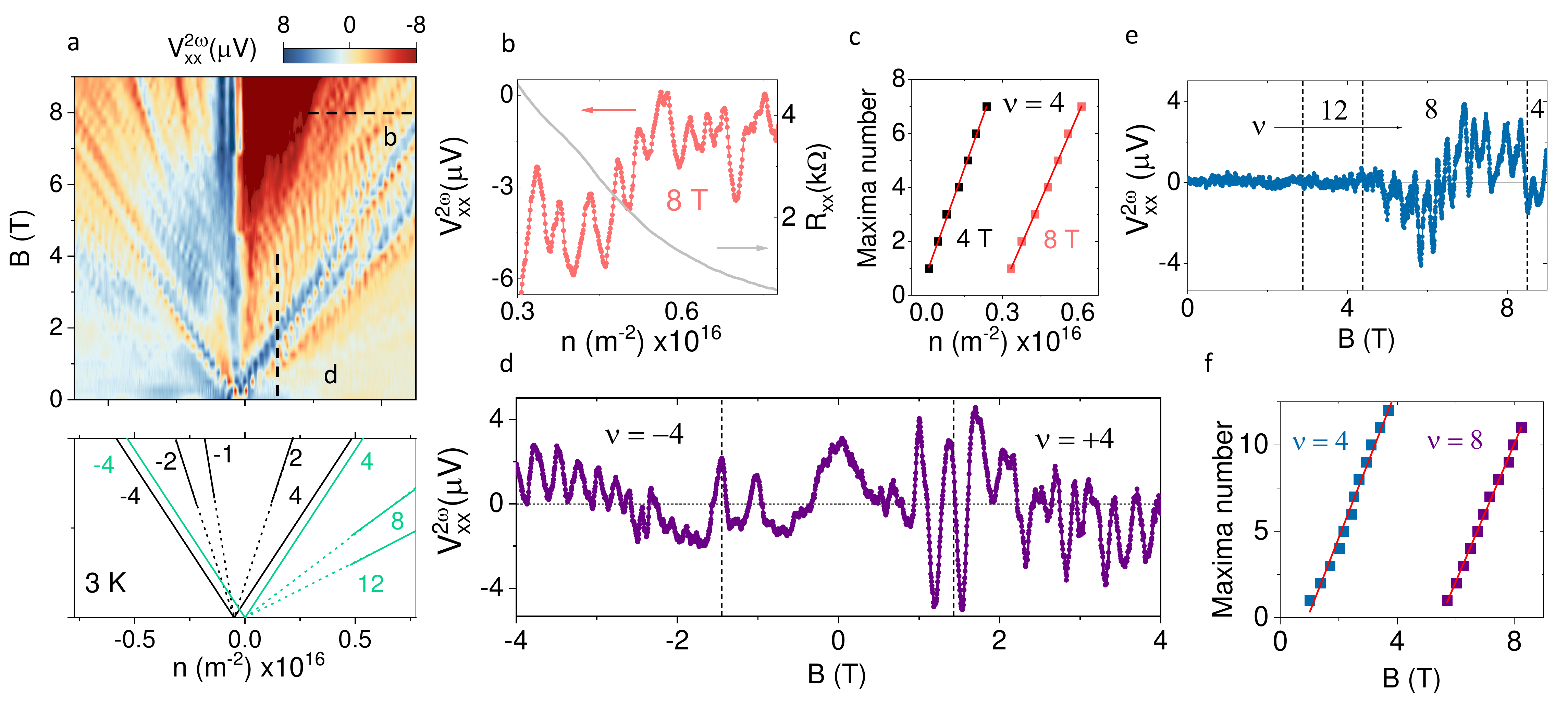}
  \captionsetup{justification=raggedright,singlelinecheck=false}
\caption{\textbf{Fabry-P\'{e}rot (FP) and Aharonov-Bohm (AhB) oscillations in magneto-thermopower:} (a) Landau fan diagram of  $V_\mr{xx}^\mr{2\omega}$. Measurements are taken at $3$~K. The bottom panel shows the reconstructed Landau level structure with filling factors ($\nu$). The green lines show the Landau fan emanating from the CNP, while the black lines show the secondary Landau fans emanating from the full-filling of the first moir\'{e} band on the hole-doping side. (b) Oscillations in $V_\mr{xx}^\mr{2\omega}$ (line cut shown by horizontal black dashed lines in Fig.~2a) at $B=8$~T, while the range of doping is confined to Landau filling $\nu=4$. The right axis shows the density-dependence of $R_\mr{xx}$ at the same $B$. (c) Doping dependence of maxima numbers in thermopower oscillations at $B=4$~T and $8$T (for same Landau filling $\nu=4$), respectively. The red lines show the linear fit. Magnetic field dependence of $V_\mr{xx}^\mr{2\omega}$ at (d) $n=1.55 \times 10^{15}$~m$^{-2}$ (vertical line-cut shown in Fig.~2a for the positive $B$ side) and (e) $n=7.75 \times 10^{15}$~m$^{-2}$. Value of $\nu$ is marked with dashed lines. (f) $B$-dependence of maxima numbers in thermopower oscillations at $\nu =4$ and $8$, respectively. The red lines show the linear fit.} 
\end{figure*}

To study the thermoelectric transport in mtBLG, we employ local Joule heating in the extended monolayer region by passing a sinusoidal current ($I_\mr{\omega}$) \cite{mahapatra2020misorientation, ghawri2020excess}. This creates a temperature gradient ($\Delta T$) along the length of the channel (left inset of Fig.~1d). The thermoelectric voltage ($V_\mr{2\omega}$) is generated in the second-harmonic ($2\omega$) which is recorded in the tBLG region of the Hall bar (xx-direction) with varying doping and heating currents. The linear response ($\Delta T \ll T$) of the measured $V_\mr{xx}^\mr{2\omega}$ is verified from $V_\mr{xx}^\mr{2\omega} \propto I_\mr{\omega}^2$ for the experimental range of heating currents (Fig.~1d). As illustrated in Fig.~1d, the doping dependence of $V_\mr{2\omega}$ shows multiple sign-reversals when the Fermi energy ($E_\mr{F}$) is varied across the low-energy band.
The sign-reversals in $V_\mr{xx}^\mr{2\omega}$ are usually attributed to changes in the quasi-particle excitations or Fermi surface topology near Lifshitz transitions \cite{jayaraman2020evidence}. The near-symmetric sign change of $V_\mr{xx}^\mr{2\omega}$ on the both side of the CNP (blue arrows in the right inset of Fig.~1d) are attributed to the positions of the van Hove singularities (vHS) in the lowest conduction/valence band of the moir\'{e} lattice \cite{ghawri2020excess}. On the hole doping side, the additional sign-reversal of $V_\mr{xx}^\mr{2\omega}$ (black arrow in the right inset of Fig.~1d) indicates the full-filling of the lowest energy valance band. Assuming filling of four electrons per moir\'{e} unit cell, we estimate $\theta \approx 0.16\degree$ from the magnitude of $n_\mr{s}$ which matches with the twist-angle estimation from Hofstadter butterfly pattern in magneto-resistance measurements. We also speculate that the observed asymmetry in doping dependence of $V_\mr{xx}^\mr{2\omega}$ originates from the particle-hole asymmetry of the band structure itself \cite{yoo2019atomic,tsim2020perfect}. 

In the degenerate regime ($T \ll T_\mr{F}$, where $T_\mr{F}$ is the Fermi temperature), the density dependence of the in-plane Seebeck coefficient can be obtained from the semiclassical Mott relation \cite{mahapatra2020misorientation},

\begin{equation}
\label{Mott relation}
S_\mr{Mott}=\frac{\pi^2 k_\mr{B}^2 T}{3|e|}\frac{1}{R_\mr{xx}}\frac{\mr{d}R_\mr{xx}}{\mr{d}V_\mr{tg}} \frac{\mr{d}V_\mr{tg}}{\mr{d}n} \frac{\mr{d}n}{\mr{d}E}\bigg\vert_{E_\mr{F}},
\end{equation}
where $\gamma=(1/R_\mr{xx})\mr{d}R_\mr{xx}/ \mr{d}V_\mr{tg}$ can be measured experimentally, and $\mr{d}n/\mr{d}E$ is the DOS ($\mr{d}V_\mr{tg}/\mr{d}n = e/C_\mr{hBN}$, where $C_\mr{hBN}$ is the known top-gate capacitance per unit area). It can be seen from Eq.~\ref{Mott relation} that the qualitative features of doping dependence of $S_\mr{Mott}$ can be captured by $\gamma$ which is compared with normalized $V^\mr{2\omega}_\mr{xx}$ in the right axis of Fig.~1f. However, we find that $\gamma$ fails to capture the multiple sign changes in $V_\mr{xx}^\mr{2\omega}$ as $\gamma$ displays sign-reversal only at CNP. Notably, the observed discrepancy of sign between $V_\mr{xx}^\mr{2\omega}$ and $\gamma$ cannot be compensated by the DOS in the Eq.~\ref{Mott relation}. We speculate that the presence of large number of vHSs in the DOS of mtBLG can give rise to large anisotropic scattering in the small parts of the Fermi surface that satisfies the Umklapp condition \cite{buhmann2013thermoelectric}. While this can lead to a violation of the Mott formula, other alternate mechanisms, for instance, correlation effects can not be entirely ruled out.

\begin{figure*}[t]
  \includegraphics[width=1.0\textwidth]{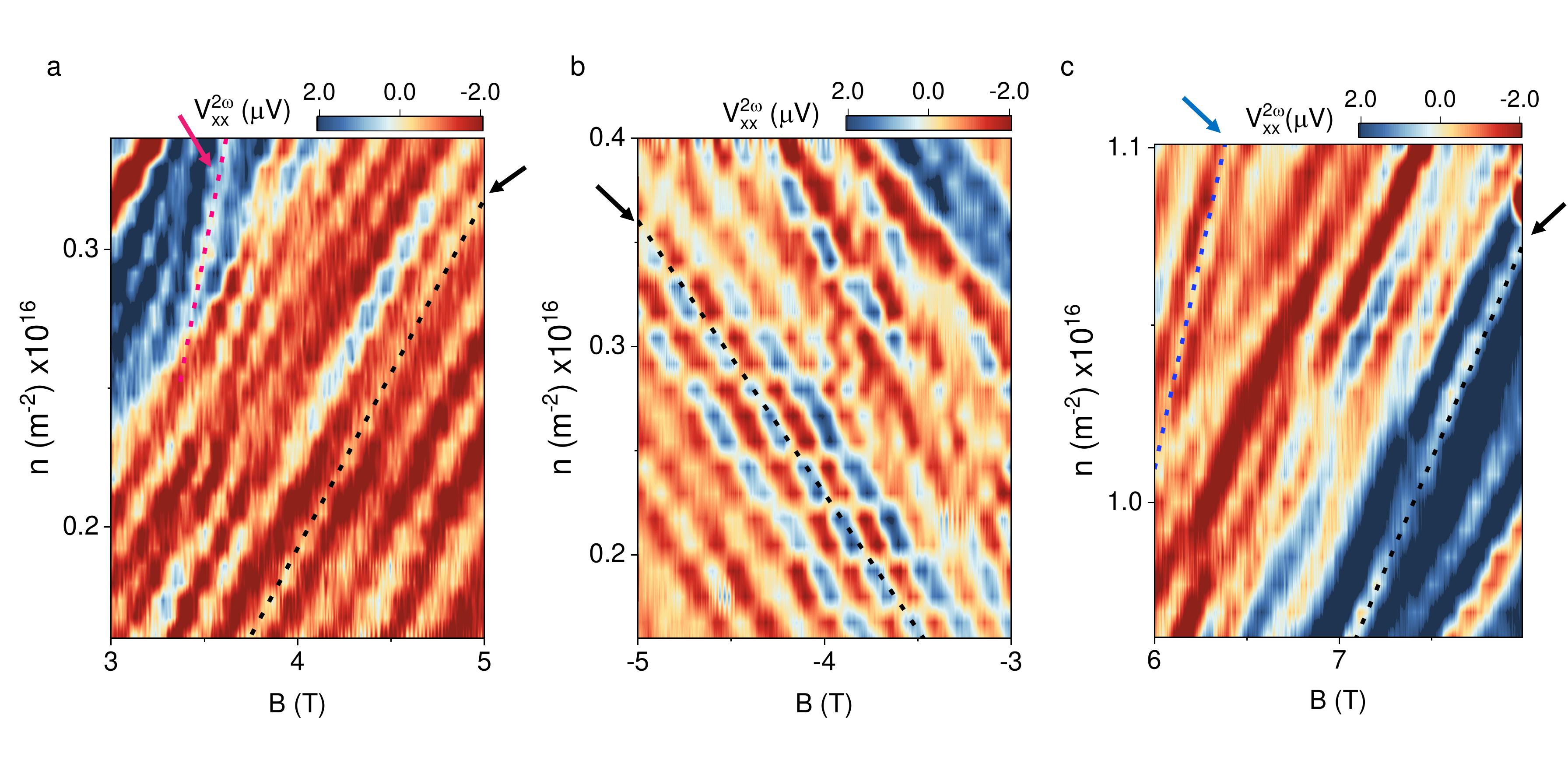}
  \captionsetup{justification=raggedright,singlelinecheck=false}
\caption{\textbf{Interference patterns in $n-B$ phase-space at $\nu=4$:} Color-plot of $V_\mr{xx}^\mr{2\omega}$ in the $n-B$ phase space at (a) $\nu=+4$, (b) $\nu=-4$, and (c) $\nu=+4$ at higher doping and magnetic field, respectively. The diagonal contrast shows the evolution of maxima/minima in interference phase in the $n-B$ phase space. The dashed lines show the magnitude of the slope $\alpha \approx 1.24 \times 10^{15}$~m$^{-2}$T$^{-1}$ (black), $2\alpha$ (blue) and $3\alpha$ (pink), respectively, as a guide to the eye.} 
\end{figure*}

We turn now to the results of the thermoelectric measurements in the same device in the presence of perpendicular magnetic field ($B$). Fig.~2a shows the Landau fan diagram of thermovoltage $V^\mr{2\omega}_\mr{xx}$ at $3$~K. $V^\mr{2\omega}_\mr{xx}$ also exhibits two sets of Landau fans emerging from the CNP and $n_\mr{s}$, with a Landau level sequence that is similar to the Hofstadter butterfly pattern in $R_\mr{xx}$. We further explore the magneto-thermopower quantitatively by varying the doping of the channel while keeping the magnetic field fixed. When the device is operated in the QH regime ($\nu=4$), $V^\mr{2\omega}_\mr{xx}$ exhibits strong oscillations as a function of $n$ as shown in Fig.~2b. However, such oscillations could not be detected in the conventional $R_\mr{xx}$ measurements in this regime (right axis in Fig.~2b). While this illustrates the sensitivity of $V^\mr{2\omega}_\mr{xx}$ compared to $R_\mr{xx}$ in QH regime and hints at the possible manisfestation of the domain walls as compressible regions which can affect heat and charge transport differently \cite{d2013thermopower}, the exact mechanism is not clear at present. We find that the oscillations in $V^\mr{2\omega}_\mr{xx}$ are periodic in $n$ (Fig.~2c). The periodicity $\Delta n \approx 4\times 10^{14}$~m$^{-2}$ does not vary appreciably when magnetic field is changed while keeping the filling factor $\nu$ the same. The periodic oscillations also persist at the higher Landau level $\nu=8$ with periodicity $\Delta n \approx 5\times 10^{14}$~m$^{-2}$, suggesting that $\Delta n$ is independent of both $\nu$ and $B$. The periodic oscillations in $V^\mr{2\omega}_\mr{xx}$ are characteristically similar to the quantum version of Fabry-P\'{e}rot (FP) interferometer for 2D systems, where the cavity is defined by the external gate assembly \cite{ronen2021aharonov,zhang2009distinct,rickhaus2018transport}. For a cavity length $L_\mr{cav}$, the interference phase in an electronic FP interferometer is given by $\phi_{\epsilon} = L_\mr{tot}k_\mr{F}/2\pi$, where $L_\mr{tot}=2L_\mr{cav}$ is the total path-difference between the two interfering trajectories. The phase $\phi_{\epsilon}$ can be modulated by the changing the energy ($E_\mr{F}$) of the injected carriers, which in turn changes the wavevector ($k_\mr{F}$) \cite{handschin2017fabry}. However, to observe the phase modulation in the presence of a Lorentz force mediated by magnetic field, the cyclotron radius has to satisfy $r_\mr{c}=\hbar k_\mr{F}/eB > L_\mr{cav}$. We estimate $r_\mr{c}\approx 9$~nm using the $2D$ density relation $k_\mr{F}=\sqrt{\pi n}$ at $8$~T, which is even smaller than $\lambda$. This implies that the $k_\mr{F} \propto \sqrt{n}$ is not valid for the charge carriers that participate in the interference, suggesting a constraint in motion for the $2$D electron system. An effective one-dimensional motion can arise if the charge carriers are already flowing in QH edge channels and remains unaffected by any further increase in the magnetic field. The absence of any periodic density oscillation in $V^\mr{2\omega}_\mr{xx}$ at low magnetic field also suggests that the trajectories only interfere in the QH regime. The valley-polarized chiral channels can be ignored as the possible origin of the thermopower-oscillations since the displacement field is not strong enough ($<0.1$~Vnm$^{-1}$) to induce a gap in the AB/BA regions, which provides the necessary topological protection \cite{san2013helical,zhang2013valley}. 

\begin{figure}[t]
  \includegraphics[clip,width=9cm]{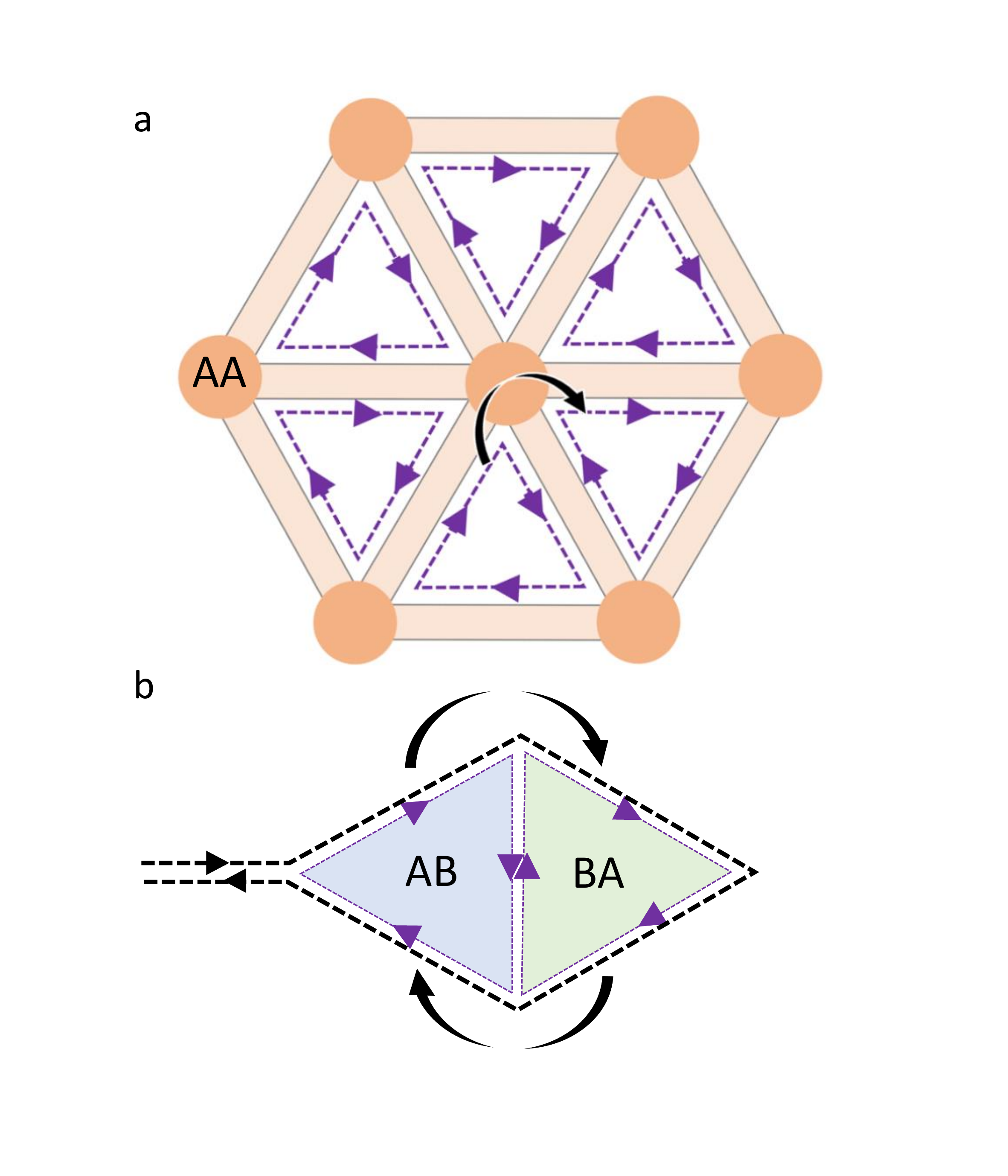}
  \captionsetup{justification=raggedright,singlelinecheck=false}
\caption{\textbf{Interference of Quantum Hall (QH) states:} (a) Schematic showing the formation of local Landau levels within the AB/BA domains in QH regime. The arrow shows tunneling or `hopping' between two neighbouring local Landau levels facilitated by the AA-stacked regions. (b) Schematic showing an example of interference between two trajectories (black dashed lines) that enclose a finite magnetic flux through a moir\'{e} unit cell. }
\end{figure}

Another modulation of the interference phase appears when we vary the magnetic field at constant doping. Fig.~2d and 2e capture the thermovoltage oscillations with varying $B$ at two different dopings. We observe the following: (1) The thermopower oscillations are periodic in $B$ (Fig.~2f). This is in stark contrast to the $1/B$-periodicity observed for Shubnikov-de Haas oscillations in $2$D systems. The $B$-linear periodicity is consistent with AhB-oscillations when the interfering trajectories enclose finite magnetic flux. We also find that the AhB-oscillations are symmetric with the direction of $B$ (Fig.~2d). This eliminates the role of contacts and the Hall component in the observed AhB-oscillations. (2) The oscillations only emerge at filling $\nu=4$ and $8$ and diminishes at higher order Landau levels. This is possibly due to weak Landau level spectrum for $\nu>8$ since the gap becomes indiscernible in both $R_\mr{xx}$ and $V^\mr{2\omega}_\mr{xx}$ in the experimental range of $n$ and $B$ (Fig.~1b and 2a).  The ubiquity of the AhB oscillations at lowest Landau levels in $n-B$ phase space indicates that the underlying mechanism of interference is profoundly connected to the formation of QH states in mtBLG. It is also important to note that the observed AhB-oscillations in the QH regime are characteristically  distinct from the AhB-oscillations in the displacement field-driven helical states of mtBLG. The latter appears even at very low magnetic fields and diminishes at large magnetic fields \cite{rickhaus2018transport}, while the former explicitly appears at the lowest Landau levels in the QH regime. For a spatially confined QH interferometer, changes in $\nu$ corresponds to changes in the Coulomb charging energy. An enhanced Coulomb interaction between the interfering Landau levels will result in an effective reduction of the area of the interferometer \cite{ofek2010role,rosenow2007influence,zhang2009distinct}. This leads to an increase in the periodicity of AhB-oscillations \cite{zhang2009distinct}. Rather surprisingly, we find that the periodicity ($\Delta B \approx 250$~mT)  of the AhB-oscillations remains independent of the filling factor $\nu$ (Fig.~2f), suggesting diminished Coulomb charging effects between the charge carrier trajectories. From the periodicity of the AB-phase, $\phi_\mr{AB} =BA_\mr{loop}/\Phi_0$, we estimate the enclosed loop area $A_\mr{loop} \approx 1.6 \times 10^{-14}$~m$^{-2}$ which is exactly twice the area of the moir\'{e} unit cell ($A_\mr{moire}$). This is remarkable since it reveals that the interference loop is much smaller than the area of the lithographically defined mtBLG channel ($\sim$~few $\mu$m$^2$), and compares with the length scale of the moir\'{e} lattice. Furthermore, we observe that the periodicity of AhB-oscillations remains unchanged ($\Delta B\approx 240$~mT) for the next nearest channel with similar $\theta$ and channel length, suggesting that $\Delta B$ depends only on the twist angle \textit{i.e,} the size of the moir\'{e} lattice (Fig.~S2, Supplementary Information, SI). This indicates that the interference area is intrinsic to the triangular domain structure of the moir\'{e} lattice across the mtBLG device, and is not related to any twist-angle inhomogeneity in the sample.

Having discussed the observation of the thermopower oscillations with varying doping and magnetic field separately, we now turn to the results of $V^\mr{2\omega}_\mr{xx}$ in the combined $n-B$ phase space. Fig.~3a and 3b illustrates the color map of $V^\mr{2\omega}_\mr{xx}$ at $|\nu|=4$. The loci of the constructive and destructive interference-phases can be identified by the contrasting diagonal lines. The cumulative index $j$ for the total interference phase can be described as

\begin{equation}
\label{Interference condition}
j=\frac{L_\mr{tot}k_\mr{F}}{2\pi}\pm \frac{A_\mr{loop}B}{\Phi_0}.
\end{equation}
The spacing between the two constructive interference is $\Delta j = \frac{L_\mr{tot}\Delta k_\mr{F}}{2\pi}\pm \frac{A_\mr{loop}\Delta B}{\Phi_0}=1$. For AhB-oscillations in mtBLG, the phase associated with path difference and the magnetic phase can add in either constructive or destructive way, which is equivalent to traversing the path in clockwise or anti-clockwise direction. This is captured by the both signs in Eq.~\ref{Interference condition} \cite{rickhaus2018transport}. However, we observe only positive slopes for the diagonal lines in $n-B$ phase space. This was further verified in the negative magnetic field at $\nu=-4$ as shown in Fig.~3b. We note that the observed positive sign of the slope is similar to the slope of the Landau fans emanating from CNP. The absence of negative slope in $n-B$ phase plane indicates that the motion of the charge carriers are allowed only in one direction, which is characteristically similar to the interference resulting from the QH states \cite{nakamura2019aharonov}. We also validate the positive slope of the constant-phase line at higher doping and magnetic field but at the same Landau level $\nu=4$ (Fig.~3c). The diagonal lines in $n-B$ phase-space at $|\nu|=4$ reveal a primary slope $\alpha \approx 1.24 \times 10^{15}$~m$^{-2}$T$^{-1}$ (indicated by the black dashed lines in Fig.~3a-c) along with weaker diagonal lines with slopes $2\alpha$ (blue) and $3\alpha$ (pink dashed lines). The latter correspond to higher harmonics of the same effective interfering area but different path length. Notably, the $n-B$ phase diagram in our device is fundamentally different compared to the $V_\mr{g}-B$ phase diagram in QH interferometers based on graphene \cite{ronen2021aharonov} or GaAs systems \cite{nakamura2019aharonov,zhang2009distinct}. In the latter case, tuning $V_\mr{g}$ will inevitably change the area of the interferometer. However, in mtBLG, tuning $V_\mr{g}$ only induces a change in $n$ and Fermi energy. This allows us to distinctly observe coupled oscillations in $n-B$ phase space without changing the interference area.

To elucidate the modulation of interference phase in $n$ and $B$ we propose the following. In the QH regime, the trajectories of the charge carriers are heavily influenced by the perpendicular magnetic field and the underlying moir\'{e} lattice. Since the domain walls separate the triangular AB/BA regions, the QH chiral channels will manifest along the domain walls (see Fig.~4a) forming local Landau levels inside the AB/BA domains \cite{bassler2021effects,kisslinger2015linear}. Importantly, the domain walls function as sample edge in the QH regime. On the other hand, the AA regions can function as nodes for the incoming and outgoing paths, which facilitate tunneling between the chiral edge channels of the neighbouring AB/BA regions. The `hopping' of the charge carriers to neighbouring QH channels gives rise to propagating modes that can carry charge and heat. FP and AhB-oscillations can arise when trajectories interfere and enclose finite magnetic flux as shown by the schematic in Fig.~4b. This is conceptually similar to the formation of propagating zigzag channels and pseudo Landau levels in mtBLG under large vertical displacement field \cite{de2020aharonov,ramires2018electrically,tsim2020perfect,fleischmann2019perfect}. However, it is important to note that the interference associated with the QH channels are qualitatively different compared to that of the valley-polarized chiral channels. The latter only appears when the Fermi energy energy is within the displacement field-induced gap, while the former depends on the local Landau level formation within the AB/BA domains. Since any AA region can host `hopping', this gives rise to the large number of the possible propagating modes, and the wave-function of the charge carrier becomes extended. This is contrary to the QH interference in GaAs systems, where the small area of the interferometer can only host finite amount of charge, and the confinement effect plays a dominant role. We believe that the extended nature of wave-function leads to diminished Coulomb repulsion between the charge carrier even though the pseudo Landau levels are localized within the domains. Furthermore, since the only the AA-stacked regions can facilitate the hopping of charge carriers, the total scattering cross section is considerably reduced. This enables phase coherence between the propagating modes over the sample dimension.
 
Remarkably, we find that the maxima/minima diagonal lines in $n-B$ phase space are linear to the variation of $n$ and $B$. This suggests that, rather importantly, when the device is operated in QH regime, $k_\mr{F} \sim n_\mr{QH}$, where $n_\mr{QH}$ is $1$D number density of charge carriers. This linear dependence is in stark contrast to the conventional $2$D case, and can be attributed the one-dimensional motion of the charge carriers.  $n_\mr{QH}$ can be estimated from the gate-induced number density ($n$) using $n_\mr{QH} = n/N_\mr{ch}$, where $N_\mr{ch}=2\sqrt{3}/\lambda$ is the density of edge channel per unit length (see SI for more details). Using $n_\mr{QH}$ in Eq.~\ref{Interference condition}, we get $L_\mr{tot}=A_\mr{loop}2\sqrt{3}g/\alpha\lambda\Phi_0$, where $g$ is the degeneracy of the carriers. The experimentally obtained $\alpha$ and $A_\mr{loop}$ renders $L_\mr{tot}\approx 450$~nm which is $\sim 4\lambda$. However, the estimated $L_\mr{tot}$ is too small to enclose an area of $A_\mr{loop} \approx 1.6 \times 10^{-14}$~m$^{-2}$ ($\approx \sqrt{3}\lambda^2$). This apparent discrepancy can be resolved when we consider the fact that all the charge carriers will be unlikely to participate in the one dimensional transport. Therefore, a fraction of charge carriers from the gate induced carrier density ($n$) should manifest in the $1D$ density $n_\mr{QH}$. This leads to an underestimation of $L_\mr{tot}$. Although the extracted area $A_\mr{loop}$ ($\approx 2A_\mr{moire}$) corresponds to a trajectory that encloses an effective area consisting of four AB/BA domains, the exact path length can manifest in many different geometries depending on the interference condition. We note that the minimum path length required to enclose $A_\mr{loop}$ is $6\lambda$ which corresponds to a triangular effective area, however, other trajectories are also possible. In the QH regime, each node (AA-stacked region) can support transfer of three incoming channels and three outgoing channels. This allows for different $L_\mr{tot}$ enclosing the same area \cite{xu2019giant}. We speculate that this leads to multiple harmonics in the observed slopes of the diagonal maxima/minima lines in $n-B$ phase space in Fig.~3a-c. 

In conclusion, we have measured magneto-thermopower transport in a mtBLG sample with $\theta \approx 0.16\degree$. We find periodic oscillations of thermovoltage in both $n$ and $B$ that is consistent with electronic FP and AhB-interference between charge carrier trajectories that enclose finite magnetic flux at low temperature ($3$~K). The periodic oscillations could only be detected in thermovoltage and not in conventional resistance measurements. The resonance patterns emerge only in the QH regime at Landau fillings $\nu=4$ and $8$, and persist throughout the different channels of the device. We estimate the enclosed loop area from the periodicity of AhB-oscillations, which is comparable to the moir\'{e} lattice size. Our observations indicate that charge carriers form local Landau levels within the AB/BA domains, and the neighbouring QH modes interfere through AA regions.  The intrinsic quantum Hall interference effect in mtBLG may act as a novel platform for realising anyonic statistics in the fractional QH regime.

\section{Materials and Methods}

\subsection{ Device fabrication}

The hBN-encapsulated mtBLG device was fabricated using the tear and stack method, where a hBN layer ($\approx 30$~nm thick) is used to tear a $\sim 40$~$\mu$m graphene layer into two parts, which were then stacked at an approximate alignment of $0.5 \degree$. The tear and stack of the mtBLG was followed by a bottom hBN pick-up and finally the heterostructure was transferred onto an oxidized silicon wafer (p$^{++}$ doped Si/SiO$_2$ wafer, with SiO$_2$ thickness of $285$~nm) at $100 \degree$~C. The dry-transfer of the hBN-encapsulated mtBLG van der Waals assembly was done using a hemispherical polydimethylsiloxane (PDMS) stamp on a glass slide coated with Poly(propylene) carbonate layer (PPC, Sigma Aldrich, CAS 25511-85-7) under a microscope with a micro-manipulator stage. The Hall-bar was patterned on the heterostructure using  electron-beam lithography followed by reactive ion etching. The electrical contacts (5 nm Cr/50 nm Au) were deposited by thermal evaporation technique in high vacuum ($\sim 10^{-6}$~mbar). An additional hBN layer($\approx 40$~nm thick) was later transferred on the patterned Hall-bar followed by e-beam lithography and metal deposition ($5$ nm Cr/$100$ nm Au) of the local top-gate over the entire mtBLG channel.

\subsection{Electrical and thermoelectric measurements}
The four-terminal resistance measurements ($R_\mr{xx}$ and $R_\mr{xy}$ were performed with \textit{ac} current excitations of $100$~nA using a low-frequency lock-in amplifier (Stanford Research System) at $226$~Hz, in a $1.5$-K pumped helium cryostat (Janis Research) in the presence of a perpendicular magnetic field. For thermoelectric measurements, local Joule heating was implemented to create a $\Delta T$ along the length of the tBLG channel. A range of sinusoidal currents ($1$-$2$~$\mu$A) at an excitation frequency $\omega$~$=$ $19$~Hz were used for Joule heating and the second harmonic thermoelectric voltage ($V_{2\omega}$) was measured using a lock-in amplifier at various channels of the device.

\bibliographystyle{naturemag}
\bibliography{Ref}

\subsection{Acknowledgement}
We gratefully acknowledge the usage of the MNCF and NNFC facilities at CeNSE, IISc. U.C. acknowledges funding from SERB through ECR/2017/001566. AG acknowledges financial support from the Department of Science and Technology (DST), Government of India. K.W. and T.T. acknowledge support from the Elemental Strategy Initiative conducted by the MEXT, Japan (Grant Number JPMXP0112101001) and JSPS KAKENHI (Grant Numbers 19H05790, 20H00354 and 21H05233).

\subsection{Author contributions}P.S.M. contributed to the device fabrication, data acquisition, analysis and data interpretation. M.G. contributed in device fabrication and assisted in data acquisition. B.G. assisted in the analysis of the data. A.J. contributed in the  data acquisition. U.C. and A.G. contributed in the data interpretation, and theoretical understanding of the manuscript. K.W. and T.T. synthesized the hBN single crystals. P.S.M, A.G. and U.C. contributed in writing the manuscript.

\subsection{Competing interests} The authors declare that they have no competing interests.
\subsection{Data and materials availability} All data needed to evaluate the conclusions in the paper
are present in the paper and/or the Supplementary Materials.

\section{Supplementary Materials}

\begin{figure}[H]
  \includegraphics[width=\columnwidth]{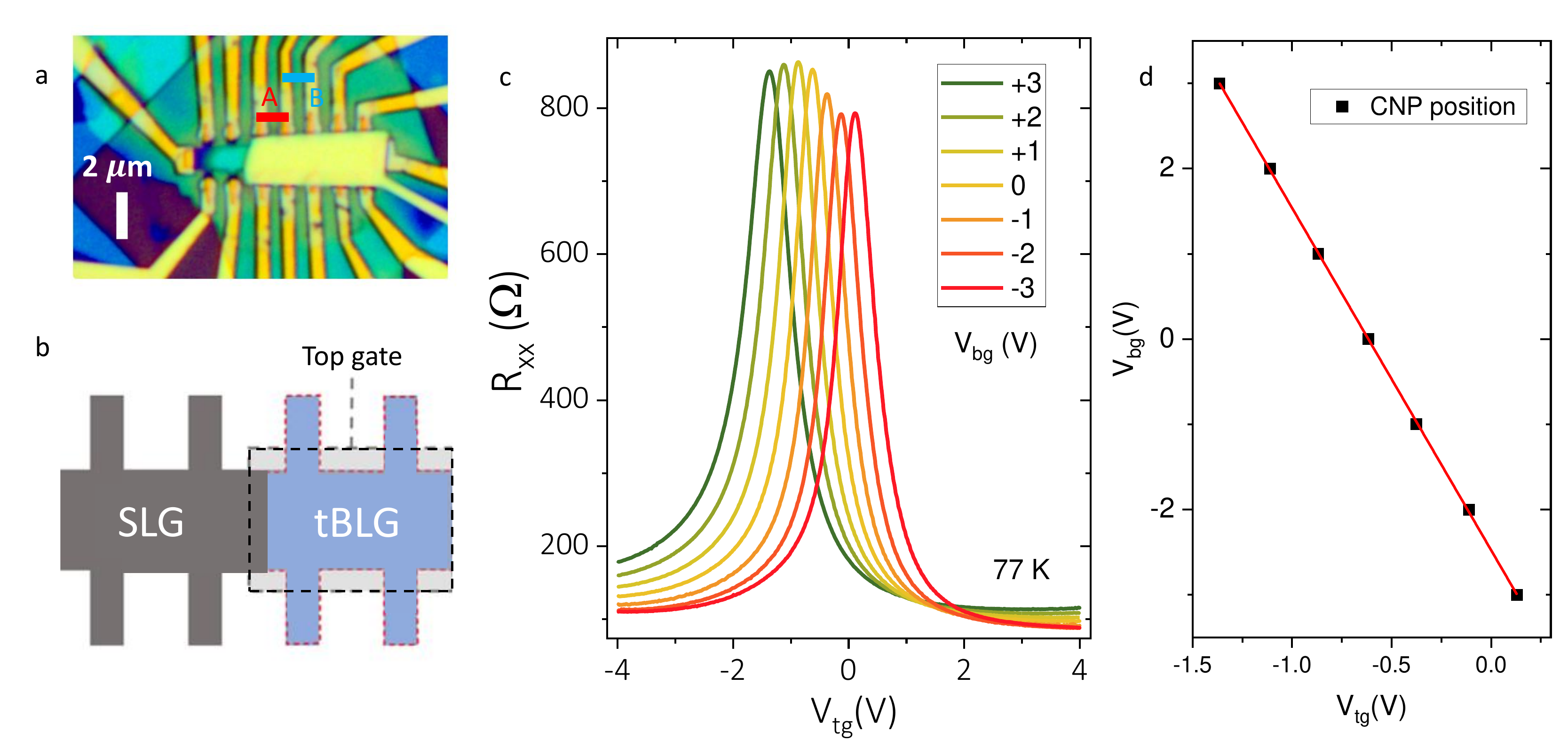}
\caption{\textbf{Device architecture:} (a) The optical image of the device. The contacts pairs are shown in red for channel A (results are described in the manuscript) and blue for channel B, respectively. (b) Schematic showing the Hall-bar geometry of the device with extended single layer graphene (SLG) part used for Joule heating in thermovoltage measurements. The top gate covers the complete area of the twisted bilayer graphene region and extends $1$~$\mu$m in length into the SLG part. (c) The variation of four-terminal longitudinal resistance ($R_\mr{xx}$) with top gate voltage ($V_\mr{tg}$) for several back gate ($V_\mr{bg}$) values at $77$~K. (d) The position of the charge neutrality point (CNP) as a function of $V_\mr{tg}$ and $V_{bg}$. The linear slope ($\approx 4$) represents the ratio of the capacitance for local top gate and global back gate. The estimated capacitance for top-gate is $\approx 0.5 \times 10^{-3}$~Fm$^{-2}$.} 
\end{figure}

\begin{figure}[H]
  \includegraphics[width=\columnwidth]{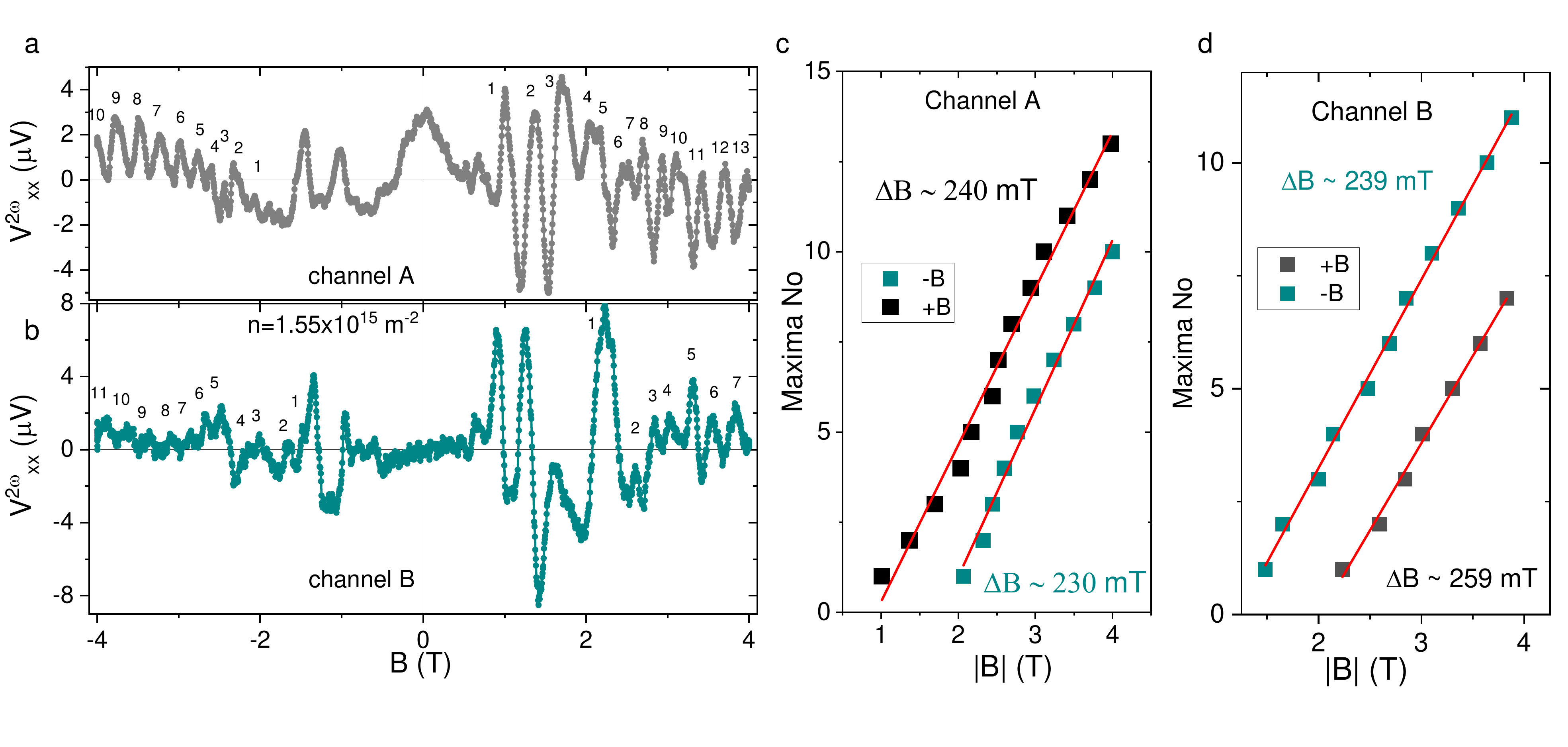}
\caption{\textbf{Aharonov-Bohm (AhB) oscillations in magneto-thermopower:} AhB-oscillations in longitudinal thermoelectric voltage $V_\mr{xx}^\mr{2\omega}$ as a function of magnetic field ($B$) for (a) channel A and (b) channel B, respectively, at a fixed number density $n=1.55 \times 10^{15}$~m$^{-2}$. The maxima positions of the AhB oscillations are marked with numbers.  $B$-dependence of maxima numbers in $V_\mr{xx}^\mr{2\omega}$  for (c) channel A and (d) channel B. The linear slopes (red lines) represent the periodicity of the AhB oscillations. } 
\end{figure}

\begin{figure}[H]
  \includegraphics[width=\columnwidth]{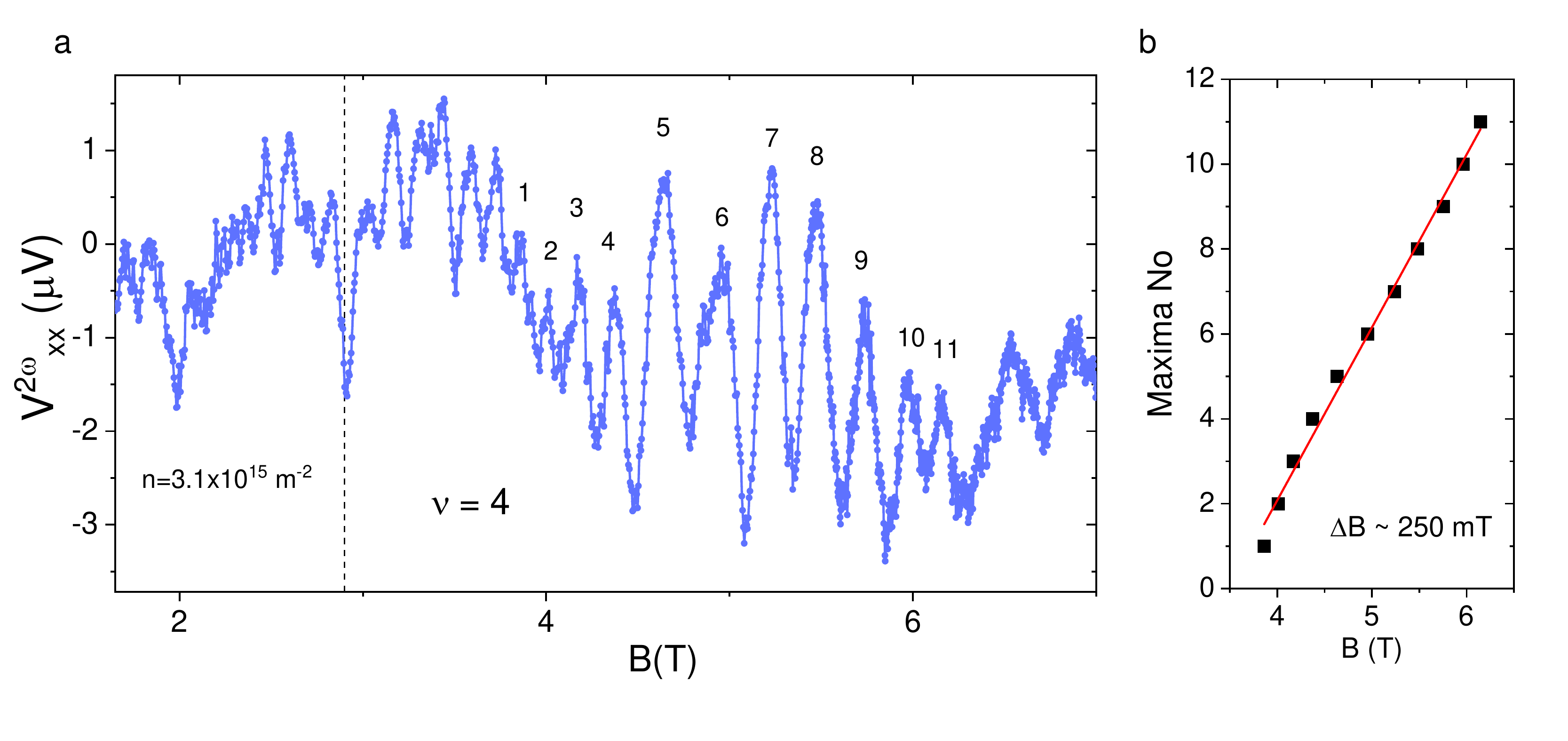}
\caption{\textbf{Aharonov-Bohm (AhB) oscillations in magneto-thermopower:} (a) AhB-oscillations in $V_\mr{xx}^\mr{2\omega}$ in channel A as a function of magnetic field ($B$) at a fixed number density $n=3.1 \times 10^{15}$~m$^{-2}$. The maxima positions of the AhB oscillations are marked with numbers. (b) $B$-dependence of maxima numbers in $V_\mr{xx}^\mr{2\omega}$. The linear slope (red line) represents the periodicity of the AhB oscillations.  }
\end{figure}

\begin{figure}[H]
  \includegraphics[width=\columnwidth]{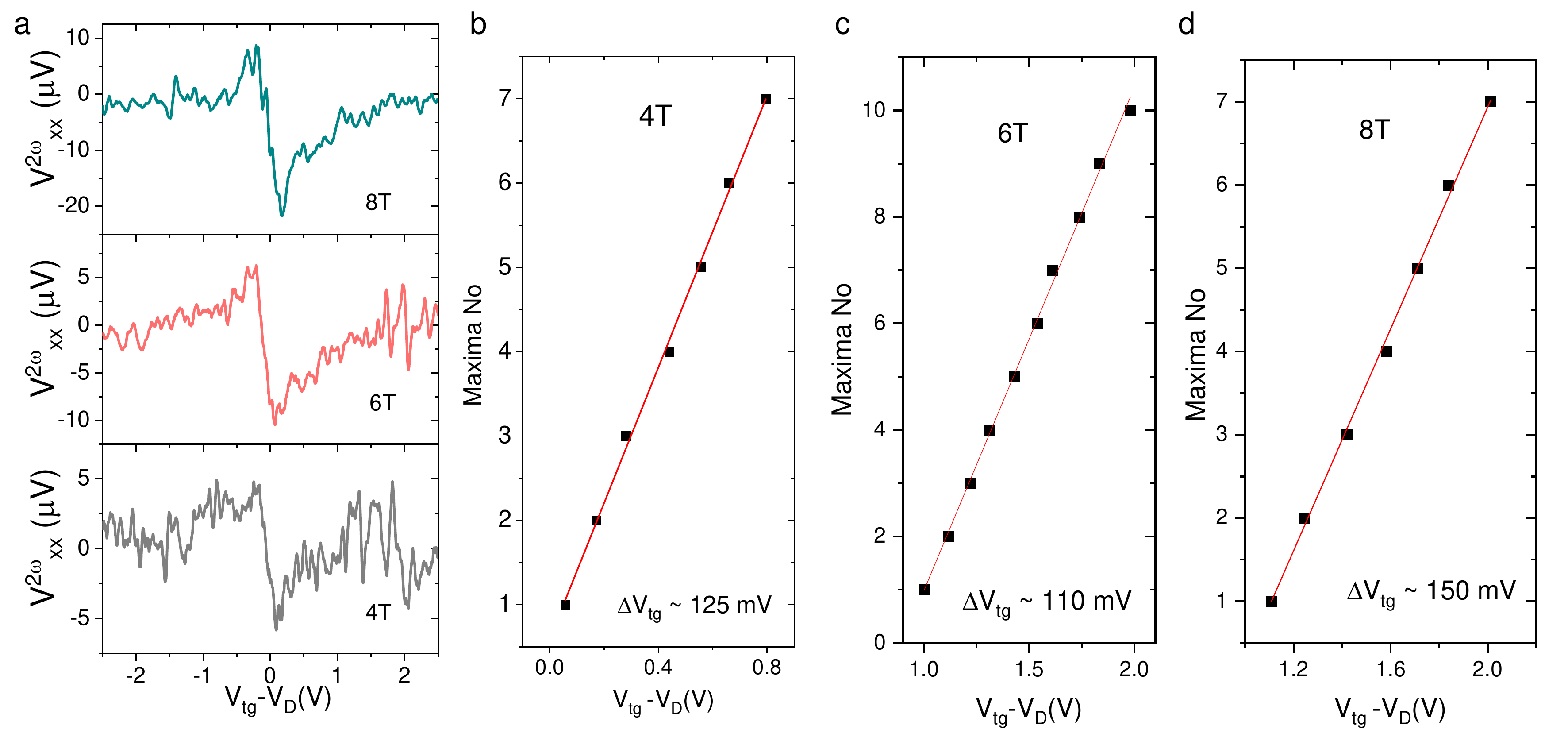}
\caption{\textbf{Fabry-P\'{e}rot (FP) oscillations in magneto-thermopower:} (a) FP oscillations of $V_\mr{xx}^\mr{2\omega}$ as a function doping ($V_\mr{tg}$) at  $B=4$~T, $6$~T, and $8$~T, respectively. Doping-dependence of maxima numbers in $V_\mr{xx}^\mr{2\omega}$  for (b) $B=4$~T, (c) $6$~T, and (d) $8$~T , respectively. The linear fits represent the periodicity of the FP oscillations. } 
\end{figure}

\begin{figure}[H]
  \includegraphics[width=\columnwidth]{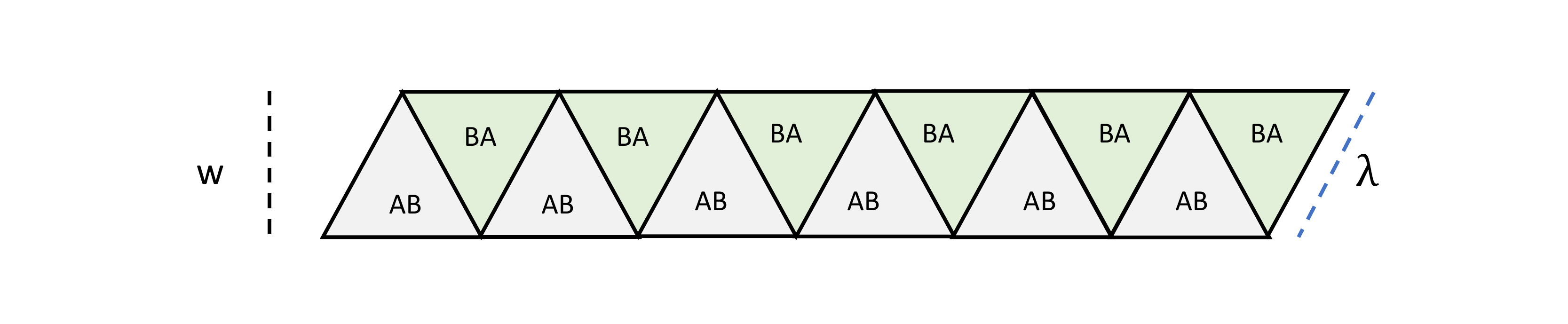}
\caption{\textbf{ Density of 1D channels in quantum Hall regime} :For a triangular network of periodicity $\lambda$ (moir\'{e} periodicity) the width ($w$) of the channel is $\sqrt{3}\lambda/2$. In one direction, the density of propagating channels are given by $N_\mr{ch}=1/w = 1/(\sqrt{3}\lambda/2)$. For three directions, $N_\mr{ch}=2\sqrt{3}/\lambda$. }
\end{figure}

 \end{document}